# Modeling Selective Intergranular Oxidation of Binary Alloys

Zhijie Xu[1,a], Dongsheng Li[1], Daniel K. Schreiber[3], Kevin M. Rosso[2], and Stephen M. Bruemmer[3]

1. Computational Mathematics Group, Fundamental and Computational Sciences Directorate, Pacific Northwest National Laboratory, Richland, WA 99352, USA

2. Physical Sciences Division, Fundamental and Computational Sciences Directorate, Pacific Northwest National Laboratory, Richland, WA 99352, USA

3. Energy and Environment Directorate, Pacific Northwest National Laboratory, Richland, WA 99352, USA

**Abstract**

Intergranular attack of alloys under hydrothermal conditions is a complex problem that depends on metal and oxygen transport kinetics via solid-state and channel-like pathways to an advancing oxidation front. Experiments reveal very different rates of intergranular attack and minor element depletion distances ahead of the oxidation front for nickel-based binary alloys depending on the minor element. For example, a significant Cr depletion up to 9 µm ahead of grain boundary crack tips were documented for Ni-5Cr binary alloy, in contrast to relatively moderate Al depletion for Ni-5Al (~100s of nm). We present a mathematical kinetics model that adapts Wagner's model for thick film growth to intergranular attack of binary alloys. The transport coefficients of elements O, Ni, Cr, and Al in bulk alloys and along grain boundaries were estimated from the literature. For planar surface oxidation, a critical concentration of the minor element can be determined from the model where the oxide of minor element becomes dominant over the major element. This generic model for simple grain boundary oxidation can predict oxidation penetration velocities and minor element depletion distances ahead of the advancing front that are comparable to experimental data. The significant distance of depletion of Cr in Ni-5Cr in contrast to the localized Al depletion in Ni-5Al can be explained by the model due to the combination of the relatively faster diffusion of Cr along the grain boundary and slower diffusion in bulk grains, relative to Al.

## I. Introduction

Preventing metal alloy failure during service life in high temperature, reactive environments remains an obstacle for new energy technologies such as light-water reactors[1,2]. It is well known that the penetration of intergranular oxidation can be strongly dependent on grain boundary characteristics and the applied stress [3,4]. For structural materials, intergranular oxidation combined with stress corrosion cracking represents a class of critical material failure mechanisms in many applications. For example, cracked structure components removed from commercial light-water reactors in service and laboratory experiment samples tested under simulated reactor environments exhibit prominent intergranular embrittlement and intergranular disintegration [5].

Understanding kinetics and mechanisms by which intergranular oxidation occurs at the atomic level is of particular interest to guide development of durable, corrosion-resistant materials and their safe application in service environments. Fast penetration of oxidants along grain boundaries is one key aspect of this broad problem. Examples are grain boundary oxidation of NiAl alloys in the temperature range 500-1000 °C[3,6] and intergranular attack along the grain boundary for Ni-Cr[7], Ni-Al[8], and similar commercial alloys (e.g. Ni-16Cr-9Fe, alloy 600)[9,10]. Figure 1 shows an optical micrograph of Ni-Al alloy after 454 hrs oxidation at 800 °C and $6\times10^{-7}$ mbar $O_2$ (Cu-Cu$_2$O stability, gaseous Rhines pack test)[3]. Under these conditions, the selective oxidation of Al occurs first at the sample surface, followed by the dissolution of oxygen into the Ni-Al solid solution with extensive ingress of oxygen along the grain boundary causing the inward growth of oxide $Al_2O_3$. Figure 2 shows images from scanning electron microscopy of the leading intergranular oxides formed in various nickel based binary alloys that were exposed to high-temperature (360 °C), hydrogenated water (Ni-NiO stability, 25 cc/kg $H_2$). Similar to the high-temperature gaseous Rhines pack exposures of Ni-Al, the Ni-5Cr and Ni-5Al alloys exhibit intergranular selective oxidation of the minor alloying specie. High resolution analyses of these oxidation fronts by transmission electron microscopy and atom probe tomography have also revealed extensive, long-range depletion of Cr from grain boundaries of Ni-5Cr for up to 9 μm beyond the oxidation front, suggesting very rapid grain boundary diffusivity[7], while grain boundary depletion in the Ni-5Al seems more localized (~100 nm ahead of the oxidation front).

The objective of this paper is to present a generic kinetics model for intergranular selective oxidation of binary alloys based on Wagner's theory for internal oxidation[11,12], and our previous study on the boundary–function method[13], oxidation kinetics[14,15] and heterogeneous reaction at the interface of two phases[16]. To model the penetration of oxidation along the grain boundary, transport processes taking place in the oxide, along the grain boundary, from bulk alloy to grain boundary, and reactions at the oxidation front itself must be considered in detail. Such transport processes should have important roles in defining the overall rate of intergranular oxidation. The same model can be applied to flat surface oxidation by neglecting mass transfer between bulk alloy and grain boundary.

The paper is organized in the following way. The kinetics model for intergranular oxidation is presented first in Section II. Application to the flat surface oxidation for Ni-Cr alloys is presented in Section III. Section IV applies the proposed model to intergranular oxidation of a set of nickel based alloys. Connections to available experiments are presented where possible, and some conclusions are drawn in Section V.

**II. Model for intergranular oxidation along a grain boundary**

The model provides governing equations describing the penetration of an oxidation front along a grain boundary in a homogeneous binary alloy *AB*. As shown in Fig. 3, the proposed model considers the diffusion of oxygen in oxide along the grain boundary (GB) to the oxidation front, the diffusion of alloy elements *A* and *B* along the grain boundary toward the oxidation front, and the diffusion of elements *A* and *B* from bulk alloy (from adjacent grains enclosing the GB under consideration), and the oxidation process at the oxide-alloy interface. Due to the consumption of oxygen and metal atoms at the oxidation front, the dynamics of the oxide-alloy interface penetrating along the grain boundary can be modeled as the transport of reactants (oxygen and alloy elements *A* and *B*) to the interface and the consumption of elements *A* and *B* at the interface due to the oxidation reaction. The model first considers the transport of all three species toward the oxidation front. The concentration profiles of oxygen (O) and elements *A* and *B* follow the standard diffusion equations,

$$\partial C_o / \partial t = D_o \nabla^2 C_o, \tag{1}$$

$$\partial C_i / \partial t = D_i \nabla^2 C_i + \left(C_i^\infty - C_i\right) R_i, \tag{2}$$

where $C_o$ is the concentration profile of oxygen in oxide along the grain boundary, and $C_i$ is the concentration profiles of alloy elements *A* and *B* along the grain boundary between two neighboring grains, with *i* denoting element *A* or *B*. $D_o$ is the effective diffusion coefficient of oxygen along the grain boundary and $D_i$ is the diffusion coefficient of element *A* or *B*. The second term on the right-hand side of Eq. (2) represents the sink (or source) term due to mass transfer between neighboring grains and the grain boundary depending on the diffusional flux of element *A* or *B* out of (or into) the grain boundary, where $C_i^\infty$ is the far-field grain boundary concentration of element *i* (can be either *A* or *B*) at equilibrium with the bulk alloy concentration. $C_i^\infty$ is assumed to be the bulk element concentration without any significant grain boundary segregation, which is reasonable for the alloys under consideration. $R_i$ is a parameter indicating the rate of mass transport between grains and the grain boundary and should be proportional to the bulk diffusion of elements $A$ and $B$. Following the original Wagner's analysis,[12] the oxidation occurs only at the oxidation front as shown in Fig. 3, where both oxygen and alloy components are consumed to produce oxide products. There is no oxygen consumption in the bulk region, and hence no oxygen consumption term in Eq. (1).

At the oxidation front, both elements $A$ and $B$ are oxidized by oxygen, where the flux of oxygen and elements $A$ and $B$ should be stoichiometrically balanced. Selecting the simplest case, by assuming the reaction at the oxidation front follows the expression

$$A^{2a+} + aO^{2-} \to AO_a, \tag{3}$$

and

$$B^{2b+} + bO^{2-} \to BO_b, \tag{4}$$

for elements $A$ and $B$, respectively, where $a$ and $b$ are stoichiometric coefficients. This mass balance at the oxidation front can be explicitly written as

$$-\frac{\alpha}{a} D_o \frac{\partial C_o}{\partial x}\bigg|^- = D_A \frac{\partial C_A}{\partial x}\bigg|^+ = k_A \left(C_A\big|^+\right)\left(C_o\big|^-\right)^a, \tag{5}$$

and

$$-\frac{1-\alpha}{b} D_o \frac{\partial C_o}{\partial x}\bigg|^- = D_B \frac{\partial C_B}{\partial x}\bigg|^+ = k_B \left(C_B\big|^+\right)\left(C_o\big|^-\right)^b, \tag{6}$$

where $k_A$ and $k_B$ are rate constants of oxidation for elements A and B, respectively. $C_i\big|^+$ and $(\partial C_i/\partial x)\big|^+$ are the element A or B concentration and concentration gradient at the interface with $|^+$ indicating the magnitude of that variable on the alloy side of the interface. $C_o\big|^-$ and $(\partial C_o/\partial x)\big|^-$ are the oxygen concentration and concentration gradient at the interface with $|^-$ indicating the magnitude on the oxide side of the interface. $\alpha$ is a parameter indicating the fraction of oxygen flux contributing to oxidize element A, and $1-\alpha$ is the fraction of oxygen flux contributing to oxidize element B. The oxide product includes both $AO_a$ and $BO_b$, with an average molar density of

$$\rho = \frac{\alpha/a + (1-\alpha)/b}{\dfrac{\alpha}{a\rho_{AO_a}} + \dfrac{1-\alpha}{b\rho_{BO_b}}}, \tag{7}$$

where $\rho_{AO_a}$ and $\rho_{BO_b}$ are the molar densities of the pure oxides $AO_a$ and $BO_b$, respectively. We also define the molar density in a binary alloy as $\rho_{AB} = C_A^\infty + C_B^\infty$, where $C_A^\infty$ and $C_B^\infty$ are the far-field concentrations of alloy elements A and B, as shown in Figure 1.

The advancing velocity of the oxidation front can be derived from the local mass conservation condition, where

$$V_s = -D_o \frac{\partial C_o}{\partial x}\bigg|^- \left(\frac{\alpha}{a\rho_{AO_a}} + \frac{1-\alpha}{b\rho_{AO_b}}\right). \tag{8}$$

Equations (1) to (8) comprise a complete set of equations that can be solved by standard interface tracking methods like phase field or level set method [17,18,19,20]. To obtain more insights into the behavior of this model, the same set of equations can be rewritten in the dimensionless form and solved approximately. By introducing the unit of length $L$, unit of time $L^2/D_o$, and the unit of velocity $U = D_o/L$, the dimensionless equations are written as:

$$\partial c_o / \partial t = \nabla^2 c_o, \tag{9}$$

$$\partial c_i / \partial t = \nabla^2 c_i / \phi_i + \left(c_i^\infty - c_i\right) r_i. \tag{10}$$

At the oxide-alloy interface,

$$-\frac{\alpha}{a} \frac{\partial c_o}{\partial x}\bigg|^- = \frac{1}{\phi_A} \frac{\partial c_A}{\partial x}\bigg|^+ = D_{aA} \left(c_A|^+\right)\left(c_o|^-\right)^a, \tag{11}$$

$$-\frac{1-\alpha}{b} \frac{\partial c_o}{\partial x}\bigg|^- = \frac{1}{\phi_B} \frac{\partial c_B}{\partial x}\bigg|^+ = D_{aB} \left(c_B|^+\right)\left(c_o|^-\right)^b, \tag{12}$$

$$v_s = -\frac{\partial c_o}{\partial x}\bigg|^- \left(\frac{\alpha}{a} w_A + \frac{1-\alpha}{b} w_B\right), \tag{13}$$

where dimensionless constants $\phi_A = D_o/D_A$, $\phi_B = D_o/D_B$ are the ratio of diffusion coefficients. The Damköhler numbers $D_{aA} = k_A L \rho_{AB}^a / D_o$, $D_{aB} = k_B L \rho_{AB}^b / D_o$ are the ratio of oxidation rate constants to the diffusion coefficients. $r_A = R_A L^2 / D_o$ and $r_B = R_B L^2 / D_o$ are the dimensionless rate constants of mass transport between bulk alloy and grain boundary. $w_A = \rho_{AB}/\rho_{AO_a}$ and $w_B = \rho_{AB}/\rho_{BO_b}$ are the ratio of molar density of alloy to that of oxide products. $v_s = V_s/U$ is the normalized interface moving velocity.

The oxygen and alloy element concentrations are normalized by $\rho_{AB}$, the molar density of the bulk alloy. Obviously the condition $c_A^\infty + c_B^\infty = 1$ is satisfied after normalization. The final solution to Eqs. (9)-(13) is only dependent on the Damköhler numbers $D_{ai}$, dimensionless constants $\phi_i$ and $r_i$, and boundary conditions ($c_o^\infty$ and $c_i^\infty$), where $i$ represents elements $A$ or $B$, respectively.

Substitution of Eq. (13) into Eqs. (11) and (12) leads to the expressions of concentration and concentration gradient of alloy elements at the oxidation front in terms of the front velocity $v_s$, the fraction parameter $\alpha$, and the oxygen concentration at the interface $c_o|^-$,

$$c_A|^+ = \frac{(\alpha/a)v_s}{\left(\frac{\alpha}{a}w_A + \frac{1-\alpha}{b}w_B\right)} \frac{\left(c_o|^-\right)^{-a}}{D_{aA}}, \tag{14}$$

$$c_B|^+ = \frac{((1-\alpha)/b)v_s}{\left(\frac{\alpha}{a}w_A + \frac{1-\alpha}{b}w_B\right)} \frac{\left(c_o|^-\right)^{-b}}{D_{aB}}, \tag{15}$$

$$\left.\frac{\partial c_A}{\partial x}\right|^+ = \frac{\phi_A v_s}{\left(\frac{\alpha}{a}w_A + \frac{1-\alpha}{b}w_B\right)} \frac{\alpha}{a}, \tag{16}$$

$$\left.\frac{\partial c_B}{\partial x}\right|^+ = \frac{\phi_B v_s}{\left(\frac{\alpha}{a}w_A + \frac{1-\alpha}{b}w_B\right)} \frac{1-\alpha}{b}. \tag{17}$$

These four equations will be used later. Numerical solutions of Eqs. (9)-(13) can only be obtained with numerical calculations such as the phase field and level set methods. Next we will try to find approximate solutions with the help of a coordinate frame moving at a velocity $v_s$ that is attached to oxidation front (as shown in Fig. 3), where $x$ is the distance away from the oxidation front. Since a one-dimensional problem is considered in this study, we assume $\partial c_j / \partial t \approx -v_s \partial c_j / \partial x$ (*j* stands for all relevant elements *O*, *A*, or *B*) at oxidation interface. This assumption hints an unchanged front shape that is not true at the very beginning. The objective here is to derive analytical solutions that can approximate the exact solutions and provide us more physical insights into this problem. Though full numerical solutions can be obtained via various advanced numerical techniques, approximate solutions of $c_o$ and $c_i$ ($i = A$ or $B$) can be found as,

$$c_o = c_o\bigg|^- - \frac{\partial c_o}{\partial x}\bigg|^- \frac{1}{v_s}\{\exp(v_s x) - 1\}, \tag{18}$$

$$c_i = c_i\bigg|^+ + \frac{\partial c_i}{\partial x}\bigg|^+ \frac{1}{\beta_i}\{1 - \exp(-\beta_i x)\}, \tag{19}$$

with appropriate boundary conditions

$$c_o(x = L_s) = c_o^\infty \text{ and } c_i(x = \infty) = c_i^\infty, \tag{20}$$

where $i$ represents element $A$ or $B$. $L_s$ is the penetration depth of oxide as shown in Fig. 3. Here $\beta_i$ is a parameter related to the characteristic length scale for an alloy element concentration profile. A large $\beta_i$ corresponds to a sharp and abrupt increase from the interface concentration $c_i|^+$ to the far-field concentration $c_i^\infty$. On the other hand, a small $\beta_i$ corresponds to a slow and smooth transition leading to an extensive depletion of element $i$. Substitution of Eq. (19) into Eq. (10) leads to the relationship

$$v_s = \frac{\beta_i}{\phi_i} - \frac{r_i}{\beta_i}, \tag{21}$$

and the length parameter $\beta_i$ can be determined as

$$\beta_i = \frac{v_s \phi_i}{2} + \sqrt{\left(\frac{v_s \phi_i}{2}\right)^2 + r_i \phi_i}. \tag{22}$$

It is now obvious that a large $\phi_i$ (slow diffusion), large $r_i$ (fast mass transfer from bulk crystalline grain to grain boundary), and a fast $v_s$ (advancing oxidation penetration) lead to large $\beta_i$ (sharp transition in concentration or localized depletion).

Substitution of expressions (14)-(17) into solutions (18), (19), and (20) leads to equations

$$c_o^\infty = c_o|^- + \frac{\exp(v_s L_s) - 1}{\left(\dfrac{\alpha}{a} w_A + \dfrac{1-\alpha}{b} w_B\right)}, \qquad (23)$$

$$c_A^\infty = \frac{\alpha/a}{\left(\dfrac{\alpha}{a} w_A + \dfrac{1-\alpha}{b} w_B\right)} \left[\frac{v_s}{D_{aA}\left(c_o|^-\right)^a} + \frac{v_s \phi_A}{\beta_A}\right], \qquad (24)$$

$$c_B^\infty = \frac{(1-\alpha)/b}{\left(\dfrac{\alpha}{a} w_A + \dfrac{1-\alpha}{b} w_B\right)} \left[\frac{v_s}{D_{aB}\left(c_o|^-\right)^b} + \frac{v_s \phi_B}{\beta_B}\right]. \qquad (25)$$

In principle, three unknowns $v_s$ (penetrating velocity of the oxidation front), $c_o|^-$ (oxygen concentration at the oxidation front) and $\alpha$ can be found from algebraic Eqs. (23), (24), and (25) for any given penetration depth $L_s$. For the sake of convenience, Eqs. (24) and (25) can be equivalently transformed into

$$\frac{c_A^\infty}{\dfrac{v_s}{D_{aA}\left(c_o|^-\right)^a} + \dfrac{v_s \phi_A}{\beta_A}} + \frac{c_B^\infty}{\dfrac{v_s}{D_{aB}\left(c_o|^-\right)^b} + \dfrac{v_s \phi_B}{\beta_B}} = \frac{\alpha/a + (1-\alpha)/b}{\left(\dfrac{\alpha}{a} w_A + \dfrac{1-\alpha}{b} w_B\right)}, \qquad (26)$$

and

$$v_s^{(b-a)} = \frac{\left[c_A^\infty \left(\dfrac{\alpha}{a} w_A + \dfrac{1-\alpha}{b} w_B\right) \dfrac{a}{\alpha} - \dfrac{v_s \phi_A}{\beta_A}\right]^b D_{aA}^b}{\left[c_B^\infty \left(\dfrac{\alpha}{a} w_A + \dfrac{1-\alpha}{b} w_B\right) \dfrac{b}{1-\alpha} - \dfrac{v_s \phi_B}{\beta_B}\right]^a D_{aB}^a}. \qquad (27)$$

Finally, by solving the coupled Eqs. (23), (26) and (27), three unknowns $v_s$, $c_o|^-$, and $\alpha$ are obtained for given stoichiometric coefficients $a$ and $b$, the set of dimensionless material parameters $D_{aA}$, $D_{aB}$, $\phi_A$, $\phi_B$, $r_A$, $r_B$, $w_A$, $w_B$ and the oxidation penetration depth $L_s$, combined with appropriate boundary conditions $c_A^\infty$, $c_B^\infty = 1 - c_A^\infty$, and $c_o^\infty$ for alloy elements $A$, $B$ and oxygen, respectively. The dimensionless molar concentration (normalized by $\rho_{AB}$) of alloy element $A$ in the oxide is $c_A|^+$ (Eq. (14)), and the molar concentration of $AO_a$ in the oxide is $\alpha/w_A$. Similarly, the molar concentration of $B$ in the oxide should be $c_B|^+$ (Eq. (15)), and the mole fraction of $BO_b$ in the oxide is $(1-\alpha)/w_B$. Therefore, the mole fraction of each element $A$, $B$ and $O$ in oxide can be found as

$$m_A = \frac{\alpha\left[1 + v_s/\left(D_{aA} c_o|^-\right)\right]}{2 + \alpha v_s/\left(D_{aA} c_o|^-\right) + (1-\alpha) v_s/\left(D_{aB} c_o|^-\right)}, \tag{28}$$

$$m_B = \frac{(1-\alpha)\left[1 + v_s/\left(D_{aB} c_o|^-\right)\right]}{2 + \alpha v_s/\left(D_{aA} c_o|^-\right) + (1-\alpha) v_s/\left(D_{aB} c_o|^-\right)}, \tag{29}$$

$$m_O = \frac{1}{2 + \alpha v_s/\left(D_{aA} c_o|^-\right) + (1-\alpha) v_s/\left(D_{aB} c_o|^-\right)}, \tag{30}$$

which satisfies the condition $m_A + m_B + m_O = 1$.

By solving $v_s$, $c_o|^-$, and $\alpha$, the concentration and concentration gradient of elements $A$, $B$, and $O$ at oxidation front can be obtained from Eqs. (14)-(17). Therefore, the concentration profiles can be obtained from Eqs. (18)-(19), and the mole concentrations of elements $A$, $B$, and $O$ in oxide can be found from Eqs. (28)-(30).

Next, we will derive solutions of coupled Eqs. (23), (26), and (27) for some simplified situations. For simplest case where the two stoichiometric coefficients $a = b = 1$ and the two molar density ratios $w_A = w_B = w$, Eqs. (23) and (26) can be simplified to

$$c_o|^- = c_o^\infty - \left[\exp(v_s L_s) - 1\right]/w, \tag{31}$$

and

$$\frac{c_A^\infty}{\dfrac{v_s}{D_{aA}(c_o|^-)} + \dfrac{v_s \phi_A}{\beta_A}} + \frac{c_B^\infty}{\dfrac{v_s}{D_{aB}(c_o|^-)} + \dfrac{v_s \phi_B}{\beta_B}} = \frac{1}{w}. \tag{32}$$

For any given oxidation penetration depth $L_s$, the oxidation penetration velocity $v_s$ can be simply solved from Eq. (32) by substitution of Eq. (31) into Eq. (32). Equation (27) can be reduced to

$$\frac{\left[c_A^\infty \dfrac{w}{\alpha} - \dfrac{v_s \phi_A}{\beta_A}\right] D_{aA}}{\left[c_B^\infty \dfrac{w}{1-\alpha} - \dfrac{v_s \phi_B}{\beta_B}\right] D_{aB}} = 1, \tag{33}$$

where fraction parameter $\alpha$ can be solved from Eq. (33) with $v_s$ solved from Eq. (32). In this way, the original equations are decoupled and significantly simplified.

## III. Flat surface oxidation of Ni-Cr alloys

For flat surface oxidation the grain boundary is not present and there is no mass transfer between the bulk alloy and the grain boundary; the two mass transport parameters are set to zero, namely $r_A = r_B = 0$. Therefore $\beta_A = v_s \phi_A$ and $\beta_B = v_s \phi_B$, and Eq. (33) can be further reduced to the simpler form

$$\frac{\left[c_A^\infty \dfrac{w}{\alpha} - 1\right] D_{aA}}{\left[c_B^\infty \dfrac{w}{1-\alpha} - 1\right] D_{aB}} = 1. \tag{34}$$

Let us now focus on the surface oxidation of Ni-Cr binary alloys. We denote Ni as the more noble element *A* and Cr as the less noble element *B*. The density of an NiO film has been reported to be $1.45 \, g/cm^3$ [21] and $1.58 \, g/cm^3$ [22]. The molar density of NiO scale is about $\rho_{NiO} = 0.033 \, mol/cm^3$ [21] and the molar density of Cr$_2$O$_3$ is about $\rho_{Cr_2O_3} = 0.03 \, mol/cm^3 \approx \rho_{NiO}$ [23]. The molar densities of pure Ni and Cr are $\rho_{Ni} = 0.15 \, mol/cm^3$ and $\rho_{Cr} = 0.14 \, mol/cm^3$. The dimensionless number $w$ in Eq. (34) can be written as,

$$w = \rho_{AB}/\rho_{AO_a} = \left[(1-c_B^\infty)\rho_{Ni} + c_B^\infty \rho_{Cr}\right]/\rho_{NiO} \approx 4.55, \tag{35}$$

where $c_B^\infty = 1 - c_A^\infty$ is simply the mole fraction of Cr in the Ni-Cr alloy. By substitution of Eq. (35) into Eq. (34), we can solve Eq. (34) to find out the dependence of $\alpha$ (oxygen fraction contributing to oxidize element $A$ or Ni in this case) on the element molar fraction in the alloy, and the ratio $r_{AB} = D_{aA}/D_{aB}$, where $D_{aA}$ and $D_{aB}$ are related to the oxidation rate constants $k_A$ and $k_B$ (For Ni-Cr, $r_{AB} \ll 1$ because Cr is less noble than Ni). Figure 4 plots the variation of fraction parameter α with the mole concentration of Cr. From Eq. (34), the parameter $\alpha$ is independent of the diffusion of two elements, and increasing with increasing ratio $r_{AB}$. For the limiting case where $r_{AB} = 0$, this dependence is reduced to a very simple relation (shown in Fig. 4)

$$\alpha = 1 - wc_B^\infty. \tag{36}$$

Element $B$ must have a critical concentration ($c_B^\infty \geq 1/w$) above which the oxidation of element $B$ will dominate over element $A$ and the oxide product will be dominated by the oxide of $B$ element. The same is also true for the other limiting case where $r_{AB} = \infty$. The penetrating velocity of oxidation can be found for a flat surface with $r_{AB} \ll 1$ from Eqs. (31) and (32), where

$$v_s \approx \frac{c_o^\infty}{\dfrac{1}{D_{aA}\left(\dfrac{1-1/w}{1/w-c_B^\infty}\right)}+\dfrac{L_s}{w}}. \tag{37}$$

Substitution of solutions (36) and (37) into (31) leads to the solution of interface concentrations for surface oxidation with $r_{AB} \ll 1$,

$$c_o\big|^- = \frac{c_o^\infty}{1+D_{aA}\left(\dfrac{1-1/w}{1/w-c_B^\infty}\right)\dfrac{L_s}{w}}, \tag{38}$$

$$c_A\big|^+ = (1-1/w), \tag{39}$$

$$c_B\big|^+ = c_B^\infty r_{AB}\left(\dfrac{1-1/w}{1/w-c_B^\infty}\right). \tag{40}$$

Obviously when one element $B$ is much less noble than the other one $A$, namely $r_{AB} \ll 1$, the interface concentration of less noble metal $B$ is proportional to the ratio $r_{AB}$. The interface concentration of oxygen is quite dependent on the parameter $D_{aA}$ ($D_{aA}$ for surface oxidation can be much larger than that of intergranular oxidation because the diffusion of oxygen is much faster in a grain boundary than that in the lattice oxide). Similarly, the mole fraction of each element in the oxide (Eqs. (28)-(30)) can be reduced to: $m_A \approx \alpha/2$, $m_B \approx (1-\alpha)/2$, and $m_O \approx 1/2$ for $r_{AB} \ll 1$ and $D_{aA} \ll 1$.

Now let us attempt to make a connection with experimental observations in the Ni-Cr alloy system. It was found that a critical concentration (critical chromium mole fraction in the alloy that is necessary to produce chromic dominated oxide scale) is about 22% with $w \approx 4.55$ (also shown in Fig. 4). This value is in good agreement with experimental findings[24,25]; the two black dots in Fig. 4 represent the data from experiment[25], where the nickel mole fraction in oxide is 0.38 (or equivalently $\alpha = 0.76$) and 0.004 ($\alpha = 0.008$) for Ni-8.0 at. % Cr alloy ($c_B^\infty = 0.08$) and Ni-21.9 at % Cr alloy ($c_B^\infty = 0.219$), respectively. It is remarkable that this simple calculation reproduces the observed existence of a critical concentration of less noble component when two alloy components have very different oxidation resistance so accurately. For alloy components with comparable oxidation resistance where $r_{AB} \sim 1$, Figure 5 presents the plot of variation of fraction parameter α with $r_{AB}$ for different concentrations of alloy element B.

**IV. Intergranular oxidation in nickel-based alloys**

In this section, we will focus on applying the proposed model to intergranular selective oxidation in Ni-based binary alloys. First, a literature survey of available bulk and grain boundary diffusion data for various alloy elements in Ni based alloys is presented in Fig. 6 and Table 1 [26-30]. Figure 6 presents the Arrhenius plot of the temperature dependence of diffusion coefficients. Table 1 presents the diffusion coefficients at temperature 360 °C that is relevant to ongoing intergranular attack and stress-corrosion cracking experiments[5,7-9,31]. These data indicate the relative relations of diffusivity for Ni, Al, and Cr with $D_{Al} \ll D_{Ni} \approx D_{Cr}$ in bulk alloys. There are less data available for Ni and Cr diffusion along grain boundaries and no reliable data was identified for Al diffusion along grain boundaries in nickel-based alloys. A reasonable approximation of the Al grain boundary diffusion coefficient (based on the data for Ni) is $3.2 \times 10^{-14}\ cm^2/s$, about seven orders of magnitude higher than its bulk diffusion.

The bulk and grain boundary diffusion data of Cr in nickel-based alloys are quite complicated. These data were provided in ref. [30] in the presence of carbon (up to 0.004% in mass) in Ni-Cr-Fe alloys. It is argued that grain boundary diffusion should increase with decreasing level of carbon content due to the chemical affinity between carbon and chromium: a common carbide forming element. Varying carbon content does not affect bulk diffusion[30,32], but significantly changes grain boundary diffusion. For Ni-5Cr alloy, we choose data in ref. [32] for Ni and Cr grain boundary diffusion, which is measured for an alloy with 80 at.% Ni and 20 at.% Cr.

The effective diffusivity of oxygen along a grain boundary was estimated to be $10^{-6}\ cm^2/s$ [33]. This number is close to (but still less than) oxygen diffusion in water $10^{-5}\ cm^2/s$ at 20°C, and is many orders of magnitude larger than that for oxygen diffusion in bulk nickel oxide.

Obviously all these diffusivity estimates would benefit from more detailed experimental and molecular simulation study, such that species transport to the advancing oxidation front through grain boundaries is better understood at a mechanistic level.

Based on the diffusivity data presented in Fig. 6 and Table 1 at given temperature of 360 °C, some model parameters can be quantitatively evaluated. In the following model calculations, the parameters used are

1) for Ni-5Cr binary alloy (*A* denotes Ni and *B* for Cr):

$D_{aA} = 10^{-3}$, $D_{aB} = 10^2$ (Cr is much less noble than Ni), $\phi_A \approx 6 \times 10^9$, $\phi_B \approx 2 \times 10^6$, $r_A \approx r_B = 10^{-7}$ (because the bulk diffusion of Ni is comparable to Cr.)

2) for Ni-5Al binary alloy (*A* denotes Ni and *B* for Al):

$D_{aA} = 10^{-3}$, $D_{aB} = 10^2$ (Al is much less noble than Ni), $\phi_A \approx 6 \times 10^9$, $\phi_B \approx 3 \times 10^7$, $r_A \approx 0.005 r_B = 10^{-7}$ (because the bulk diffusion of Ni is much slower than Al).

The solubility of oxidizing species (H$_2$O) in SiO$_2$ was estimated to be $5 \times 10^{-5}\, mol/cm^3$ [34]. Therefore, we estimate the boundary condition (concentration of oxygen specie) $c_o^\infty \approx 1 \times 10^{-4}$ for oxygen diffusing through the oxide in an aqueous environment. Boundary conditions for $c_A$ and $c_B$ are determined from the atomic fraction of each element in bulk alloys as $c_A^\infty = 0.95$ and $c_B^\infty = 0.05$.

Given this set of parameters, solutions of Eqs. (31)-(33) are presented in Figs. 7-11 for both Ni-5Cr and Ni-5Al binary alloys. Figure 7 plots the variation of oxidation penetration velocity $v_s$ with penetration depth $L_s$. There is slight decrease of the penetration speed with increasing penetration depth because of the increasing length of transport for oxygen to the progressing oxidation front. With the first 100 µm of oxidation penetration, the velocity $v_s$ is about $5 \times 10^{-8}\, mm/s$ for Ni-5Cr and $7 \times 10^{-8}\, mm/s$ for Ni-5Al respectively. This oxidation penetration speed is about one order of magnitude faster than $2 \times 10^{-9}\, mm/s$, than that of surface oxidation of pure Ni at 500 °C [21]. Furthermore, if one draws a comparison to measured crack growth rates under stress, our model intergranular oxidation rates are within an order of magnitude of stress crack corrosion crack growth rates measured to be ~9×10$^{-8}$ for Ni-5Cr and 2×10$^{-7}$ for Ni-5Al under chemically similar conditions (20% cold-worked, 20 MPa/m, 330ºC) [35]. Because our model presently neglects stress effects, this suggests a possible correlation between the SCC crack growth rate and the grain boundary oxidation penetration speed.

The oxygen concentration profiles are presented in Fig. 8. It was found that the oxygen concentration within the oxide is almost constant and independent of the penetration depth for both alloys, due to the relatively fast penetration of oxygen along the grain boundary[3].

The concentration profiles for Ni are shown in Fig. 9. Figure 10 shows the concentration profile for Cr and Al. There is not much difference for Ni concentration between both alloys in Fig. 9. However, significant depletion of Cr ahead of the oxidation front is found in contrast to the much more localized depletion of Al, in general agreement with experimental observations of very long-range Cr depletion in Ni-5Cr[7]. Fast diffusion of Cr along grain boundaries seems responsible for this extended depletion. The relatively slow diffusion of Al combined with fast mass transfer from grain to grain boundary explains the localized Al depletion. Quantitative calculation gives $\beta_{Cr} = 1.2$ and $\beta_{Al} = 36.6$, or the characteristic length of depletion of 0.8 µm (up to 9 µm observed in experiment) for Cr and 0.03 µm for Al (~100s of nm observed in experiment), respectively. The atomic compositions of each element in alloy (right) and oxide phases (left) are shown in Figure 11.

By varying the value of mass transfer coefficient $r_A$ for both alloys, and if grain boundary migration can be ignored, the effect of mass transport from grain to grain boundary can be studied quantitatively. In such a case it was found that an increase of mass flow into the grain boundary will increase the oxidation penetration velocity for both alloys, as shown in Fig. 12. A more pronounced increase for Ni-5Al alloy was observed than that of Ni-5Cr alloy because of the relatively faster bulk diffusion of Al. For the same reason, a larger decrease of α, the oxygen fraction taken by Ni, with increasing $r_A$ can be expected for Ni-5Al alloy, as shown in Fig. 13. It was shown that the dimensionless number $\beta_i$ is increasing (or equivalently decreasing in the distance of depletion) with increasing $r_A$ in Fig. 14. The mass transport from neighboring grains to grain boundary compensates the depletion of Al and Cr along the grain boundary (oxidation front acts as the sink term for intergranular diffusion) and decreases the distance of depletion. The atomic compositions of Al and Cr in metal oxide also increase with $r_A$ as shown in Fig. 15. In reality, however, grain boundary migration can complicate or obfuscate this simplified picture by modifying the effective mass flow of the minor element into the grain boundary.

## IV. Conclusions

A mathematical kinetics model is presented to describe the intergranular oxidation of binary alloys. Transport of species both along the grain boundary and transverse to it through adjacent grains are considered. Approximate solutions were obtained for this moving interface problem in a similar semi-analytical way for stationary boundary[13]. Model parameters can be reasonably determined from the literature data of transport coefficients of relevant elements or from atomistic simulations. For planar surface oxidation, the model predicts a critical concentration of minor element (Cr in this case) where the oxide of the minor element (chromium oxide) becomes dominant over the oxide of the major element. For grain boundary oxidation, the model provides a quantitative relationship between oxidation penetration velocity, the penetration depth, and the depletion distance ahead of the advancing front, based on microscopic transport properties of species. The extended depletion distance of Cr and relatively localized depletion of Al that has been observed experimentally can be explained by the fast diffusion of Cr along the grain boundary and the slow Cr diffusion in bulk grains, in comparison to Al. The model provides a simple framework for understanding transport kinetics and pathway dependence of intergranular attack, and is capable of atomistic parameterization based on explicitly evaluated microscopic diffusivities.

**Acknowledgments:**

Funding for this work was provided by the U.S. Department of Energy (DOE) Office of Science, Office of Basic Energy Sciences, Materials Science and Engineering Division. Pacific Northwest National Laboratory (PNNL) is operated by Battelle for DOE under Contract DE-AC05-76RL01830.

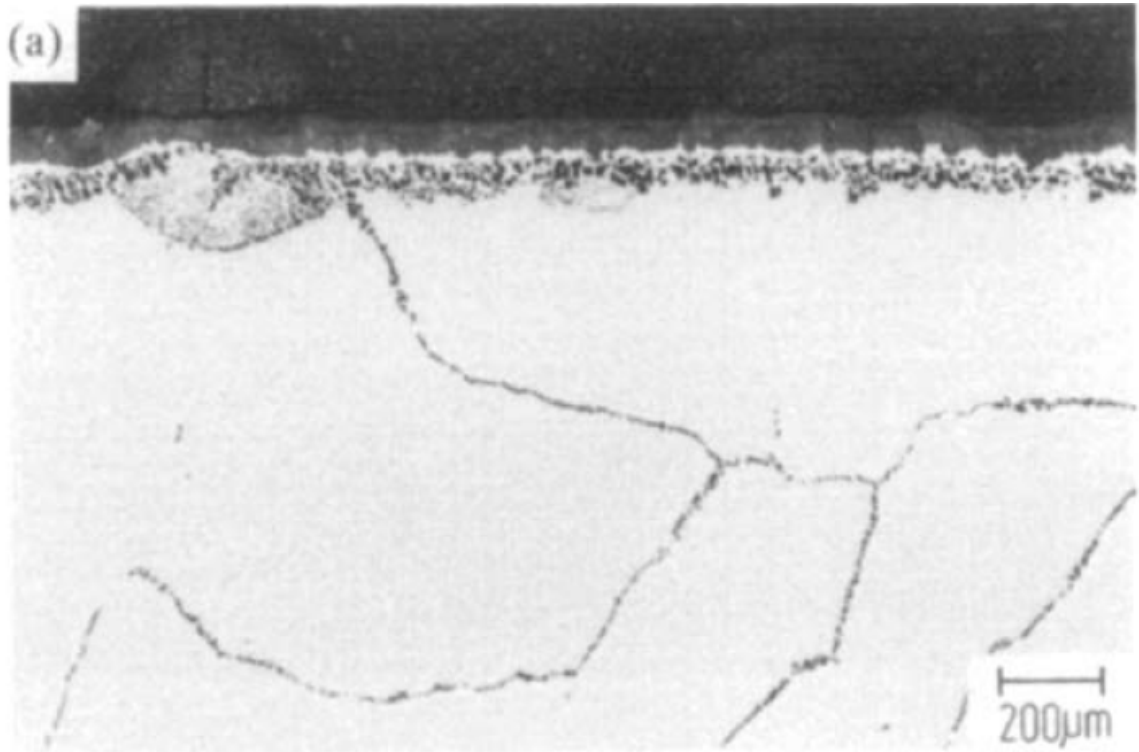

Figure 1. Optical micrograph of a cross section of Ni-50Al after 454 hours oxidation at 800C showing the intergranular oxidation. (Reprinted from Corrosion Science Vol. 36, pp. 37-53, "The oxidation of NiAl-III. Internal and intergranular oxidation" by M.W. Brumm, H.J. Grabke, B. Wagemann, Copyright (1994), with permission from Elsevier)

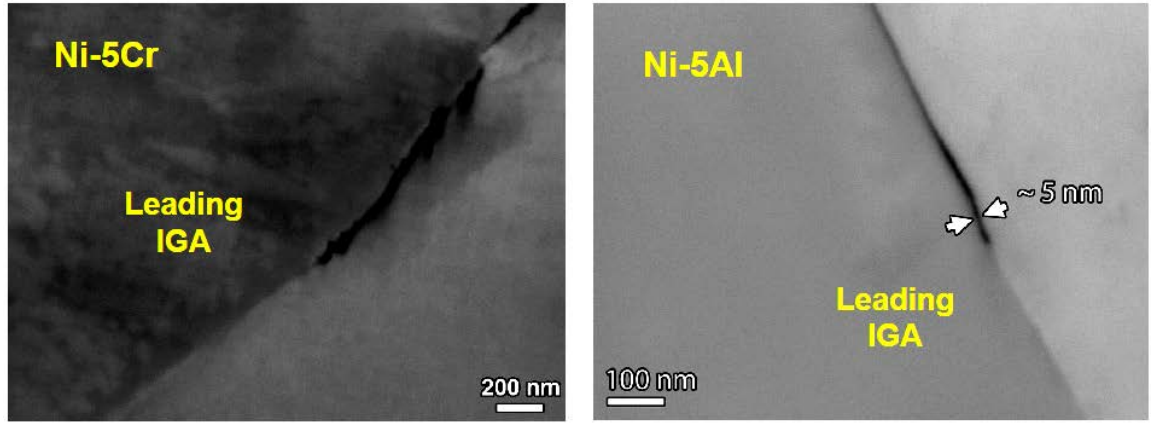

Figure 2. Leading intergranular attack for Ni-5Cr and Ni-5Al formed exposure to high-temperature, hydrogenated water.

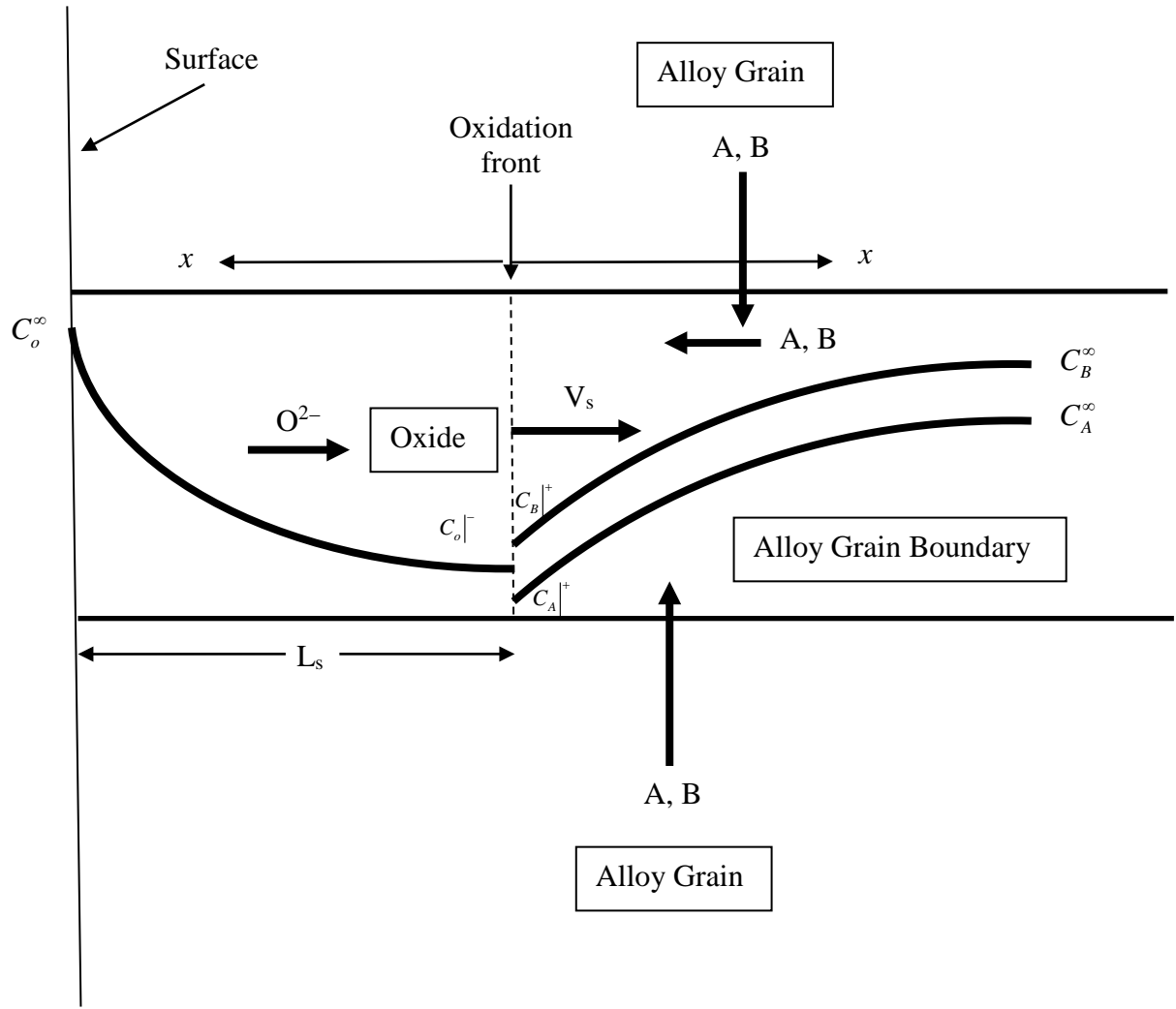

Figure 3. Schematic plot of intergranular oxidation of a model AB binary alloy.

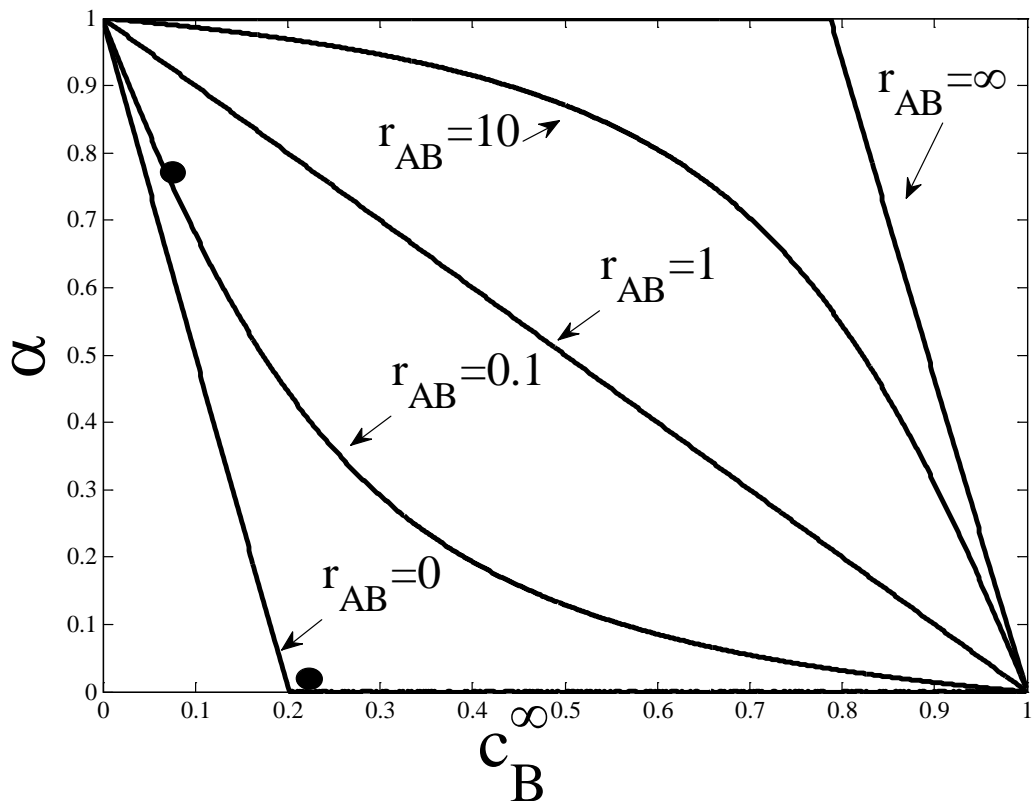

Figure 4. Dependence of α (oxygen fraction taken by element A) on the atomic composition of element B for various ratio of the oxidation rate constants between two alloy components.

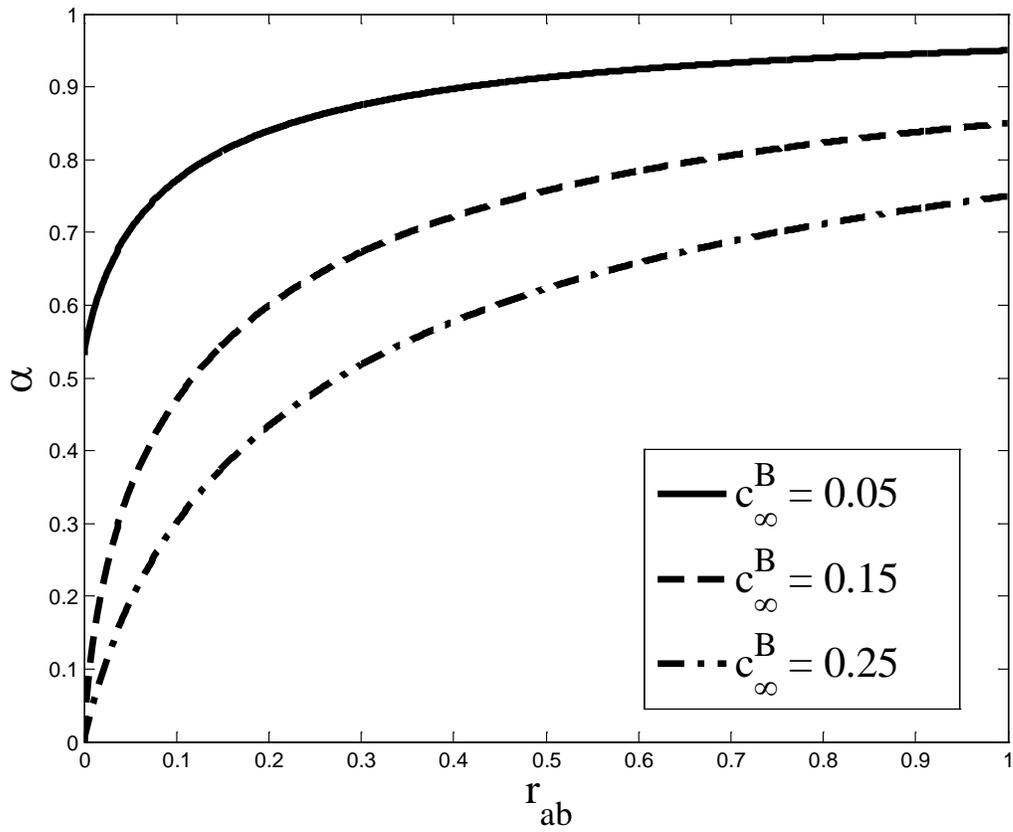

Figure 5. Dependence of α (oxygen fraction taken by element A) on the ratio of the oxidation rate constants between two alloy components for various alloy element B concentration.

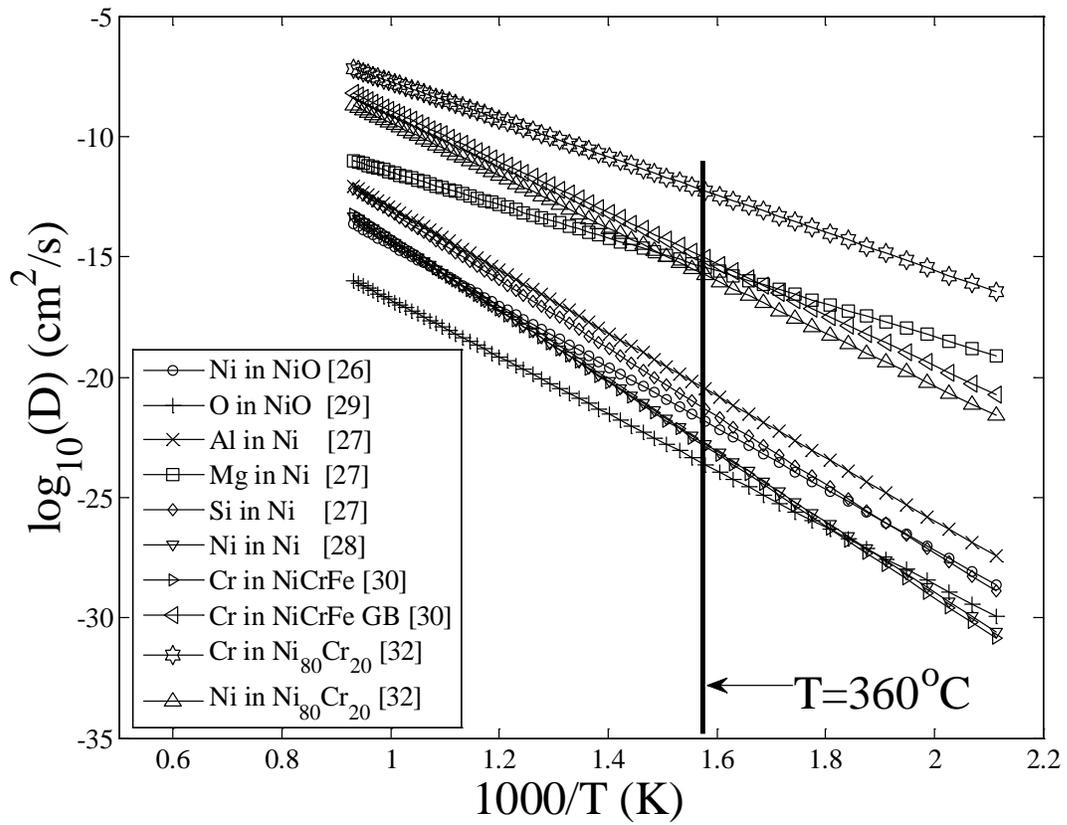

Figure 6. Arrhenius plot of the temperature dependence of diffusion coefficients from literatures.

|  | Bulk | Grain Boundary |
|---|---|---|
| O in NiO | $2.3 \times 10^{-24}$ [29] | $10^{-6}$ [33] |
| Ni in NiO | $1.4 \times 10^{-22}$ [26] |  |
| Ni in Ni | $1.7 \times 10^{-23}$ [28] |  |
| Al in Ni | $3.2 \times 10^{-21}$ [27] | $3.2 \times 10^{-14}$ |
| Cr in NiCrFe | $1.3 \times 10^{-23}$ [30] | $8.6 \times 10^{-16}$ [30] |
| Ni in $Ni_{80}Cr_{20}$ |  | $1.7 \times 10^{-16}$ [32] |
| Cr in $Ni_{80}Cr_{20}$ |  | $5.5 \times 10^{-13}$ [32] |

Table 1. Bulk and grain boundary diffusion coefficients (cm$^2$/s) at 360°C for Nickel-based alloys taken from the literature.

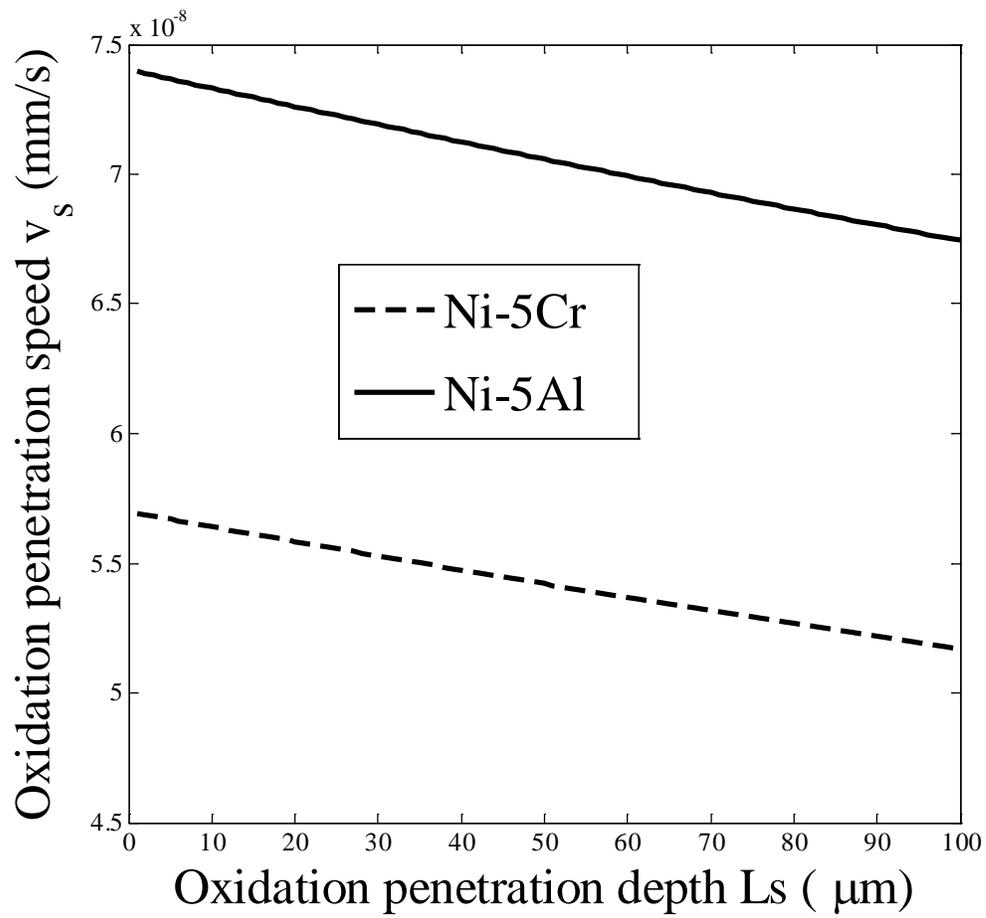

Figure 7. The variation of oxidation penetration velocity with the penetration depth Ls.

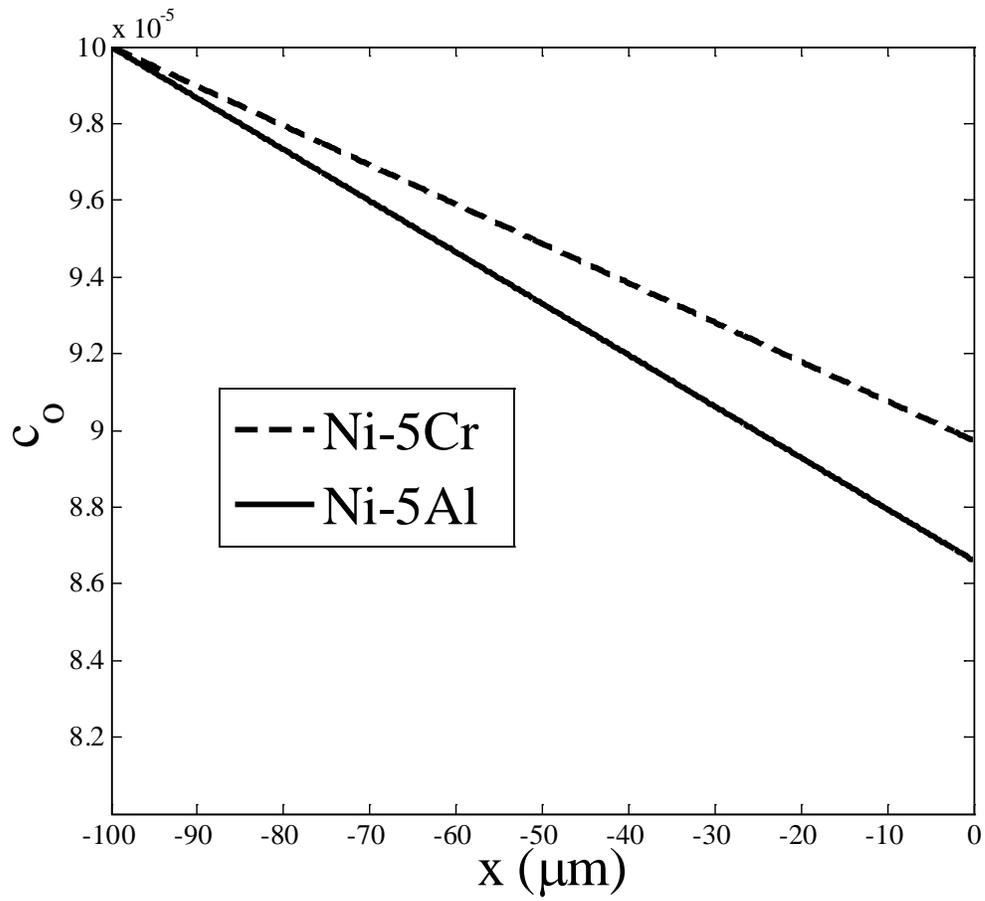

Figure 8. The spatial variation of oxygen concentration when the penetration depth

Ls=100μm

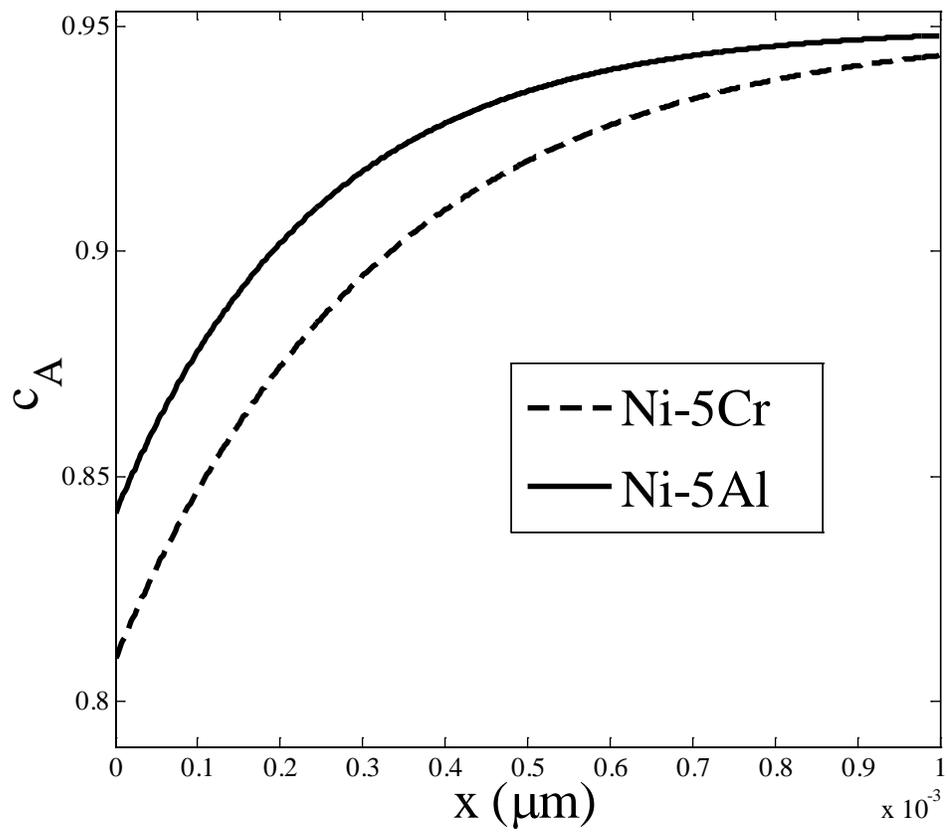

Figure 9. The spatial variation of Ni concentration when the penetration depth Ls=100μm

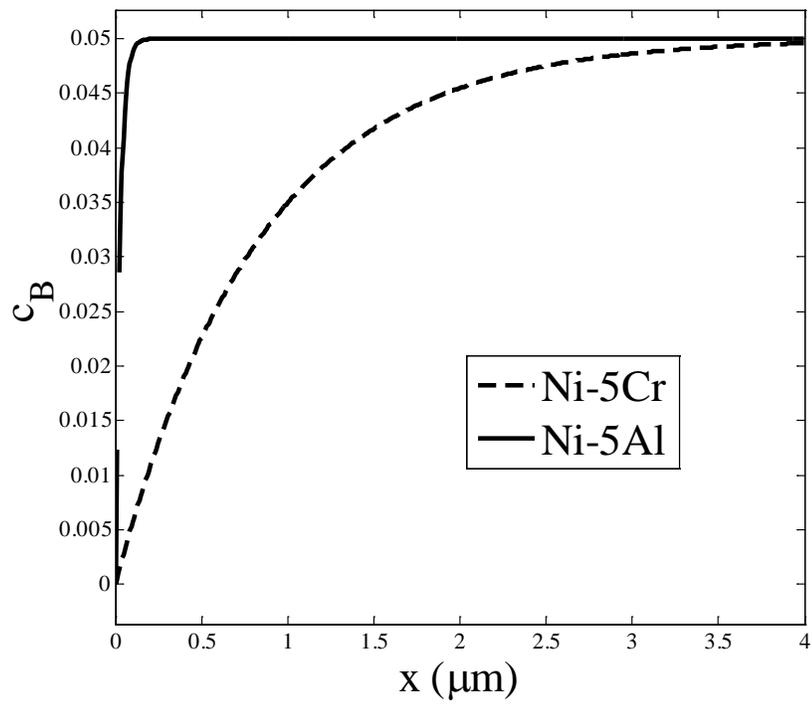

Figure 10. The spatial variation of Al (Solid line) and Cr (dash line) concentrations when the

penetration depth Ls=100μm

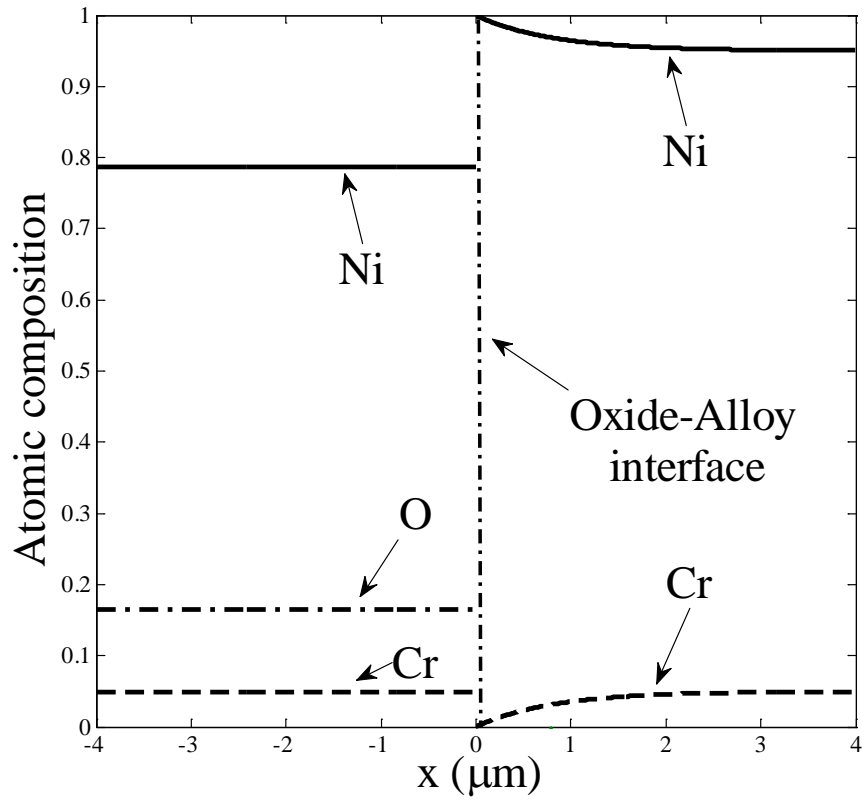

Figure 11. The spatial variation of atomic compositions for elements Ni (solid line), Cr (dash line) and O (dash dot line) when the penetration depth Ls=100μm

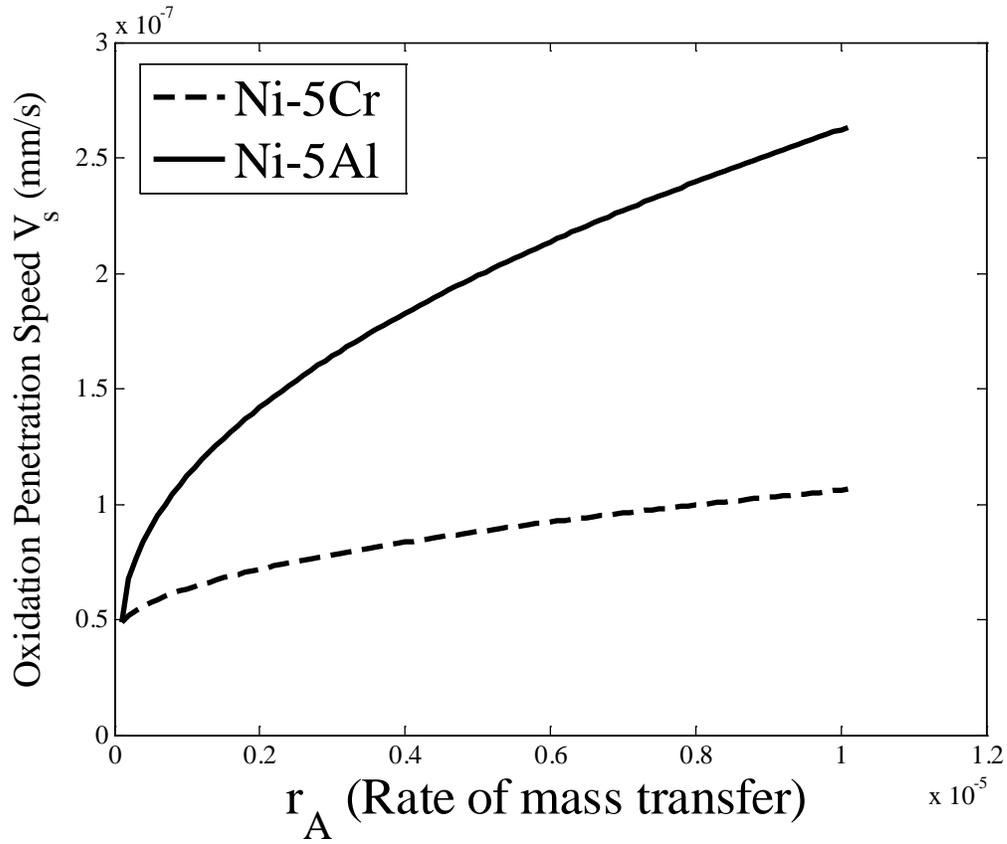

Figure 12. The variation of Vs (oxidation penetration velocity) with $r_A$ (rate of mass transfer) for Ni-5Al (solid line) and Ni-5Cr (dash line)

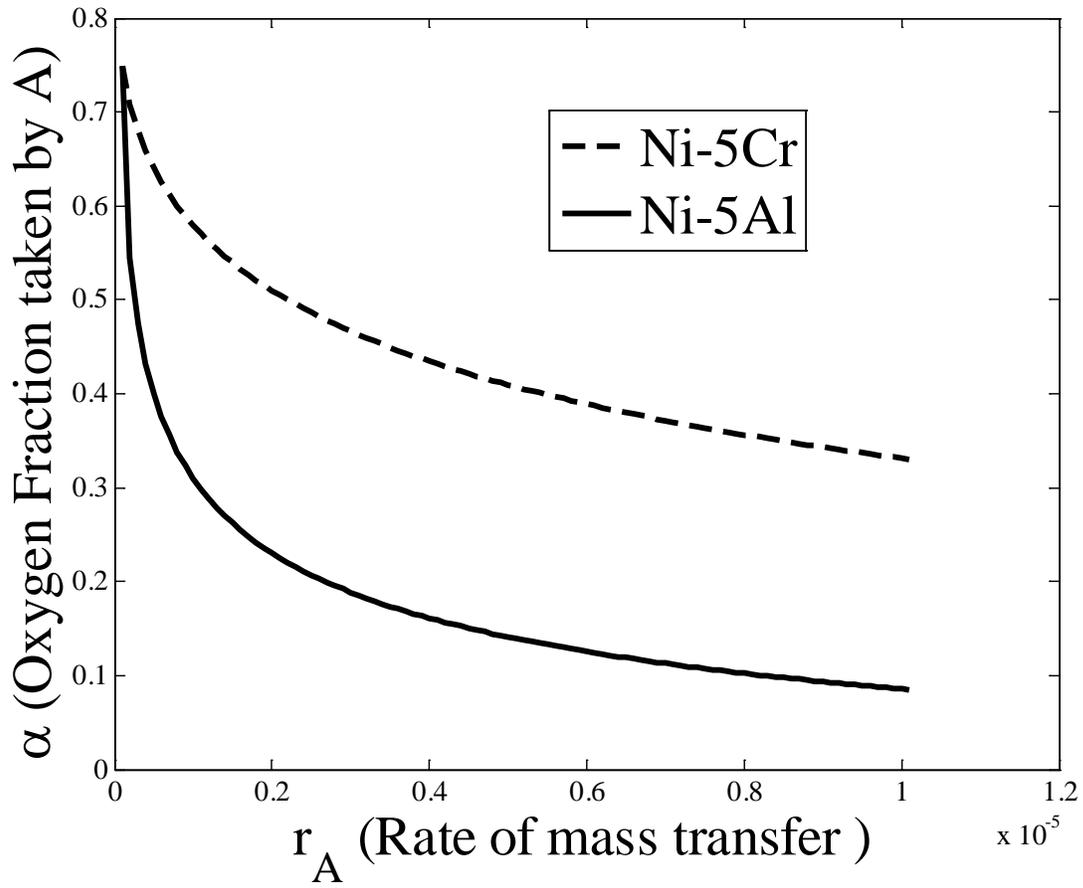

Figure 13. The variation of Vs (oxidation penetration velocity) with $r_A$ (rate of mass transfer)

for Ni-5Al (solid line) and Ni-5Cr (dash line)

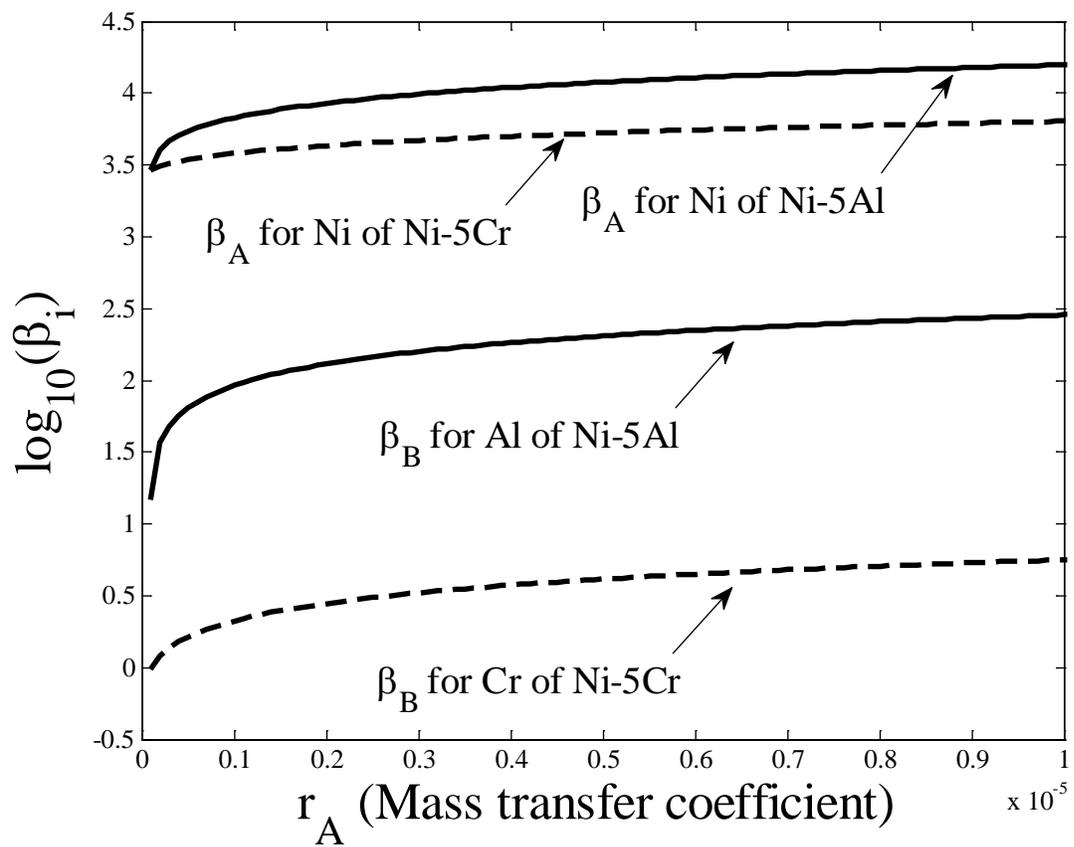

Figure 14. The variation of β$_i$ (characteristic length of concentration) with r$_A$ (rate of mass transfer) for Ni-5Al (solid line) and Ni-5Cr (dash line)

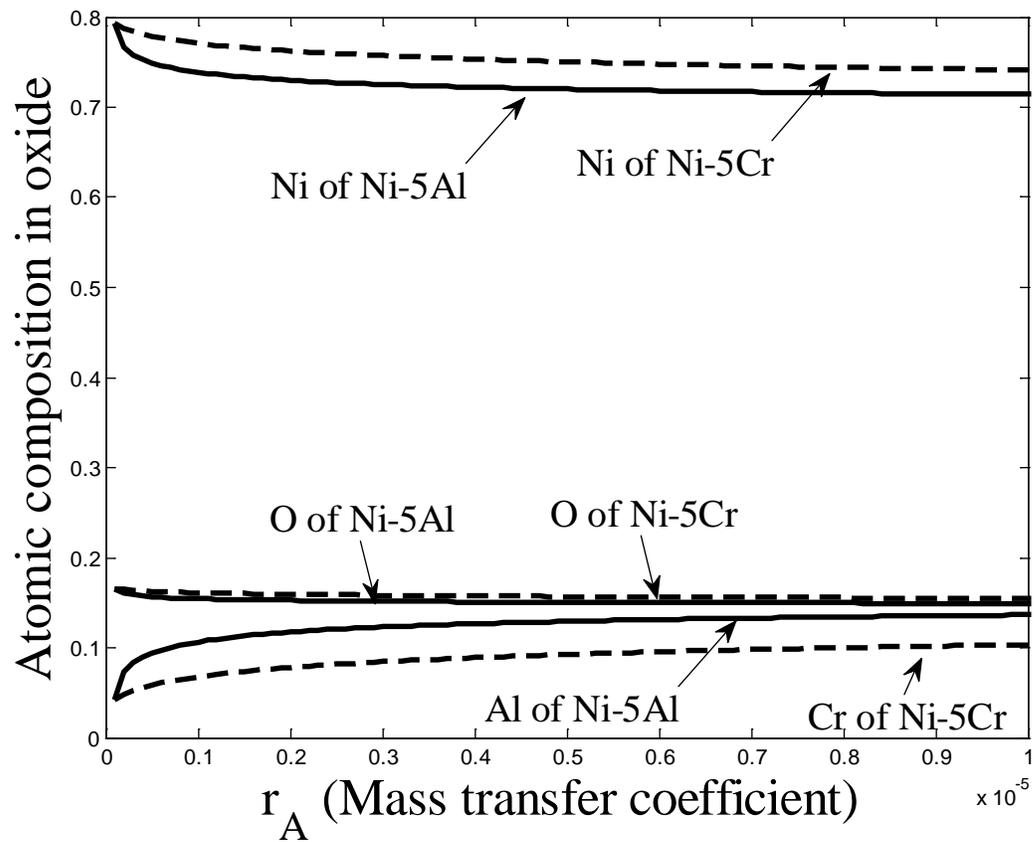

Figure 15. The variation of atomic composition in oxide for O, Cr, Al, Ni with $r_A$ (rate of mass transfer) for Ni-5Al (solid line) and Ni-5Cr (dash line).